\documentclass[aps,prl,groupaddress,twocolumn,nopacs,letterpaper,nobibnotes]{revtex4}

\PassOptionsToPackage{breaklinks}{hyperref}

\usepackage{graphicx}
\usepackage{amsmath}
\usepackage{color}
\usepackage{ulem}
\usepackage[dvipsnames]{xcolor}
\usepackage[colorlinks=true,urlcolor=blue,citecolor=blue,linkcolor=blue]{hyperref}

\renewcommand{\k}{\vec{k}}

\newcommand{\crea}[1]{#1^{\dag}}
\newcommand{\ani}[1]{#1^{\vphantom{\dag}}}
\newcommand{\up}{\uparrow}
\newcommand{\down}{\downarrow}
\newcommand{\ave}[1]{\left\langle#1\right\rangle}
\renewcommand{\vec}[1]{\boldsymbol{#1}}

\begin{document}

\title{Insulator-pseudogap crossover in the Lieb lattice} 
\author{Kukka-Emilia Huhtinen}
\author{P\"{a}ivi T\"{o}rm\"{a}}
\email{paivi.torma@aalto.fi}

\affiliation{Department of Applied Physics, Aalto University, 00076
  Aalto, Finland}
  
\begin{abstract}
We study the attractive Hubbard model in the Lieb lattice to
understand the normal state above the superconducting critical
temperature in a flat band system. We use cluster dynamical mean-field
theory to compute the two-particle susceptibilities with full quantum
fluctuations included in the cluster. At interaction strengths lower
than the hopping amplitude, we find that the normal state on the flat
band is an insulator. The insulating behavior stems from the
localization properties of the flat band. A flat-band enhanced
pseudogap with a depleted spectral function arises at larger
interactions. 
\end{abstract}

\maketitle

 Flat, dispersionless energy bands host superconductivity governed by
 the quantum geometry and topology of the
 band~\cite{Peotta2015,Julku2016,Liang2017,hazra:2019,xie:2019}. The
 predicted transition temperature exceeds exponentially that of
 conventional
 superconductors~\cite{Khodel1990,Khodel1994,Heikkila2011,Kopnin2011},
 promising superconductivity at elevated temperatures. The
 observations of superconductivity and insulating phases at quasi-flat
 bands in twisted bilayer
 graphene~\cite{Cao2018,Cao2018b,Yankowitz2019} reinforce such
 prospects. The nature of the normal state above the critical
 temperature of a flat band superconductor is, however, an outstanding
 open question. A Fermi liquid is excluded due to the absence of a
 Fermi surface~\cite{Gurin2001}. As the band width and kinetic energy
 are zero, any attractive interaction could be anticipated to cause
 pairing already in the normal state. Indeed preformed
 pairs~\cite{Tovmasyan:2018,Torma:2018} and a
 pseudogap~\cite{Swain2020,Hofmann2019} have been suggested. Here we
 show that the normal state in a Lieb lattice features, for decreasing
 interaction, a \textit{crossover from a flat-band enhanced pseudogap
   to an insulator}. For small interactions, when lowering the
 temperature, one thus expects an insulator-superconductor transition
 unique to flat band systems.

The mean-field superconducting order parameter vanishes at the
critical temperature. Understanding the normal state is thus
inherently a beyond mean-field problem. Two-particle properties, that
is, four-operator correlations must be evaluated with quantum
fluctuations included. For this, we use a cluster expansion of
dynamical mean-field theory (DMFT)~\cite{Maier2005}. DMFT has been
used previously to investigate for example the normal state properties
of the attractive single-band Hubbard
model~\cite{Capone2002,Keller2001,Toschi2005}, and its cluster
variants for studies of pairing fluctuations with different
geometries~\cite{Chen2015}. The normal state of (partially or nearly)
flat band 
systems with repulsive interactions has also been studied in
Refs.~\cite{Shinaoka2015,Kumar2019,Kumar2020,Huang2019,Sayyad2020}. We
calculate 
the orbital-resolved pair, spin and charge susceptibilities based on
two-particle Green's functions.

\begin{figure}
  \centering
  \includegraphics[width=\columnwidth]{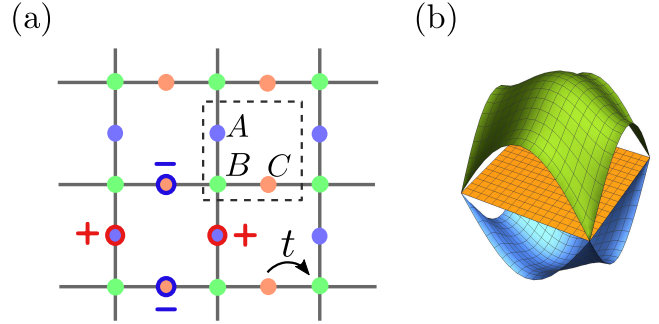}
  \caption{(a) The Lieb lattice. The unit cell indicated with dashed
    lines is used as a cluster in DMFT. The three sublattices are
    labeled $A$, $B$ and $C$ as shown. An example of a localized flat
  band state is shown in the lower left-hand corner. The $A$ and $C$
  sites on a square plaquette have amplitudes of alternating
  sign. Since we consider only nearest-neighbor hopping, this state is
  localized by destructive interference. (b) Band structure of the
  Lieb lattice. The flat band is at energy $E=0$. On the
  dispersive bands, the saddle points at the edge of the Brillouin
  zone at energies $E=\pm 2$ lead to Van Hove singularities where the
  density of states also diverges.}
  \label{fig.lieb}
\end{figure}

We focus on the Lieb lattice, shown in Fig.~\ref{fig.lieb}a, due to
its conceptual simplicity and experimental relevance. The localized
flat band states reside at the lattice sites $A$ and $C$ only, and can
be 
monitored separately from site $B$ both in experiments and
simulations. This gives a means of distinguishing flat band effects.
The Lieb lattice has been realized experimentally for ultracold
gases~\cite{Taie2015,Ozawa2017}, in designer lattices made by
atomistic control~\cite{Slot2017,Drost2017} and in photonic
lattices~\cite{Mukherjee2015}. Some covalent-organic frameworks are
also predicted to provide the Lieb lattice~\cite{Cui2020}. We relate
our predictions of the insulator and pseudogap phases to generic flat
band effects to unveil their relevance beyond the Lieb lattice, for
instance for moir\'e materials~\cite{Balents2020} where bands
of different degree of flatness can be designed.

The band structure of the Lieb lattice features two dispersive bands
and a perfectly dispersionless flat band, 
\begin{equation}
E_{\pm}(\k) = \pm t \sqrt{2} \sqrt{2 + \cos(k_x) + \cos(k_y)}, \;
E_{\rm FB} = 0.
\end{equation}
where the indices $+$ and $-$ refer to the upper and lower dispersive
band, respectively, FB refers to the flat band and $t$ is the
nearest-neighbor hopping 
amplitude. The lattice constant is taken as $a=1$. The band structure
is shown in Fig.~\ref{fig.lieb}(b). The flat band in the Lieb lattice
is related to states localized due to destructive interference. An
example of such a state is pictured in
Fig.~\ref{fig.lieb}(a). Importantly, particles on the flat band occupy
only the A and C sublattices.  

We study the attractive Hubbard model with the Hamiltonian
\begin{align}
  H &= \sum_{\sigma}\sum_{i\alpha, j\beta}
  t_{ij} \crea{c}_{\sigma, i\alpha} \ani{c}_{\sigma, j\beta} -
  \sum_{\sigma}\sum_{{\rm i}\alpha}\mu_{\sigma} n_{\sigma,i\alpha} \\
  &+
  U\sum_{i\alpha}(n_{\up,i\alpha} -1/2)
  (n_{\down,i\alpha} -1/2),
\end{align}
where $\crea{c}_{\sigma, i\alpha}$ is the creation operator for a
fermion with spin $\sigma=\up,\down$ in the unit cell $i$ and the
sublattice $\alpha=A,B,C$ and $n_{\sigma,
  i\alpha}=\crea{c}_{\sigma,i\alpha}\ani{c}_{i\alpha}$. The hopping
amplitude $t_{ij}$ is $1$ between connected sites and $0$ otherwise;
below, 
all energies are thus in units of the hopping $t$. The 
on-site interaction strength is denoted by $U$.

To study normal state properties, we use a cluster expansions of
DMFT where we use one unit cell of the
Lieb lattice as our cluster (see Fig.~\ref{fig.lieb}a). In this
method, the lattice model is mapped to an effective Anderson impurity
model, and the lattice quantities are computed self-consistently. The
self-energy is assumed to be uniform and local to each unit cell,
$\Sigma_{ij}\approx \Sigma\delta_{ij}$. Here, $\Sigma$ is a matrix in
the orbital indices. 

More precisely, the principle of dynamical mean-field theory is as
follows. The self-energy $\Sigma(i\omega_n)$ and Green's
function $G(\vec{k},i\omega_n)$ are related by the Dyson equation
\begin{equation}
  G(\vec{k},i\omega_n) =
  [G^0(\vec{k},i\omega_n)^{-1}-\Sigma(i\omega_n)]^{-1},
  \label{eq.lattice_dyson}
\end{equation}
where $G^0(\vec{k},i\omega_n)$ is the non-interacting Green's
function and $\omega_n$ are fermionic Matsubara frequencies. To map
the lattice model to an impurity model, we consider the local 
part of the Green's function
$\overline{G}(i\omega_n)=\sum_{\k}G(i\omega_n,\vec{k})$. The bath
Green's function of the 
impurity model is obtained from
\begin{equation}
  \mathcal{G}^0(i\omega_n)
  = [\overline{G}^{-1}(i\omega_n) + \Sigma(i\omega_n)]^{-1}.
  \end{equation}
In this work, the impurity problem defined by
$\mathcal{G}^0(i\omega_n)$ is solved using an interaction expansion
continuous time Monte Carlo solver
(CT-INT)~\cite{Assaad2007,Gull2011}. The solution of the impurity
problem  provides the impurity Green's function
$\mathcal{G}(i\omega_n)$ and a new estimate for the self-energy. In
DMFT, the self-energy of the impurity is equal to the self-energy of
the lattice, so the result can be plugged into
Eq.~\eqref{eq.lattice_dyson}. This procedure is repeated until the
self-consistency condition $\overline{G}(i\omega_n) =
\mathcal{G}(i\omega_n)$ is fulfilled.  

Calculation of the two-particle susceptibilities is a central and
highly non-trivial part of our work. The main ideas are discussed
here, and further details are given in the supplementary
material~\cite{supplementary}. This procedure~\cite{Rohringer2012}
allows to 
compute the generalized susceptibilities
\begin{equation}
  \chi_{ijkl}(\tau_1,\tau_2,\tau_3) =
  G^{(4)}_{ijkl}(\tau_1,\tau_2,\tau_3) -
  G_{ij}(\tau_1,\tau_2)G_{kl}(\tau_3,0),
\end{equation}
where $G^{(4),{\rm ph}}_{ijkl}(\tau_1,\tau_2,\tau_3) =
  \ave{T_{\tau}[\crea{c}_{i}(\tau_1) \ani{c}_{j}(\tau_2)
      \crea{c}_{k}(\tau_3) \ani{c}_{l}(0)]}$ is the two-particle
  Green's function. Here, $T_{\tau}$ is the imaginary time ordering
  operator and $\tau_i$ are imaginary times. The indices $i,j,k,l$
  contain the spin and the orbital indices $A$, $B$ or $C$. To
  conveniently 
  define the spin, charge and pairing susceptibilities, we define the
  Fourier transform in the particle-hole (ph) and particle-particle
  (pp) channels as follows:
  \begin{align}
  \chi_{ijkl}^{{\rm ph},\omega,\omega',\nu} =&
  \int_0^{\beta}\int_0^{\beta}\int_0^{\beta} {\rm d}\tau_1{\rm
    d}\tau_2{\rm d}\tau_3 \nonumber\\
  &\chi_{ijkl}(\tau_1,\tau_2,\tau_3)
  e^{-i\omega\tau_1} e^{i(\nu+\omega)\tau_2}
  e^{-i(\nu+\omega')\tau_3},\\
  \chi_{ijkl}^{{\rm pp},\omega,\omega',\nu} =&
  \int_0^{\beta}\int_0^{\beta}\int_0^{\beta} {\rm d}\tau_1{\rm
    d}\tau_2{\rm d}\tau_3 \nonumber\\
  &\chi_{ijkl}(\tau_1,\tau_2,\tau_3)
  e^{-i\omega\tau_1} e^{i(\nu-\omega')\tau_2} e^{-i(\nu-\omega)\tau_3},
  \end{align}
  where $\nu$ is a bosonic Matsubara frequency and $\omega$ and
  $\omega'$ are fermionic Matsubara frequencies.
Both $\chi^{\rm pp}$ and $\chi^{\rm ph}$ contain the same
information. 

The generalized susceptibilities can be computed with the impurity
solver for the cluster. However, within DMFT, the local cluster
susceptibilities are not equal to the lattice
susceptibilities. Instead, the self-consistency is only at the level
of the local irreducible vertex function $\Gamma$, which is the
two-particle equivalent to the self-energy. Like the
self-energy, the 
irreducible vertex is assumed to be momentum-independent. It is
related to the generalized susceptibilities by the Bethe-Salpeter
equation
\begin{align}
  &\chi^{{\rm c},\omega,\omega',\nu}_{ijkl} =
  \chi_{0,ijkl}^{{\rm c},\omega,\omega',\nu} \nonumber\\
  &+
  \chi_{0,ij'ji'}^{{\rm 
      c},\omega,\omega'',\nu}\Gamma_{i'j'k'l'}^{{\rm 
      c},\omega'',\omega''',\nu}\chi_{k'kl'l}^{{\rm
      c},\omega''',\omega',\nu}.
\end{align}
Here, c denotes the channel and $\chi_0^{{\rm c}}$ is the bare
susceptibility, for example in the particle-hole channel
$\chi_{0,ijkl}^{{\rm ph},\omega,\omega',\nu}=
-\beta\delta_{\omega,\omega'}
G_{il}(i\omega)G_{kj}(i(\omega+\nu))$. Repeated indices are summed
over. The Bethe-Salpeter equation can
be written separately for the local cluster quantities and the lattice
quantities. The irreducible vertex in each channel is obtained by
inverting the Bethe-Salpeter equation for the impurity
quantities. Within DMFT, the irreducible vertex of the impurity is
equal to that of the lattice, so the lattice susceptibilities can then
be computed by plugging the result in the Bethe-Salpeter equation for
the lattice quantities.

\begin{figure}
  \centering
  \includegraphics[width=\columnwidth]{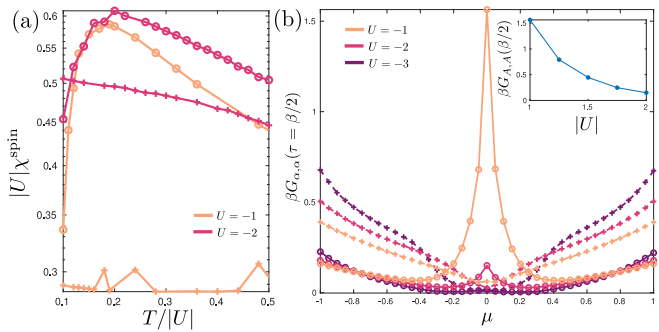}
  \caption{(a) Orbital-resolved spin susceptibilities
    $\chi_{\alpha}^{\rm spin}$ as a function of $T/|U|$ at
    half-filling for interaction strengths $U=-1$ and 
    $U=-2$. The susceptibilities are multiplied by the interaction
    strength for visual clarity. In both (a) and (b),
    results for the $A$ site are plotted with circles and results for
    the $B$ site with crosses. The flat band states are located only
    on the $A/C$ sites. (b) The Green's function $\beta
    G_{\alpha,\alpha}(\beta/2)$ at inverse temperature $\beta=20$ as a
    function of chemical potential $\mu$. Pairing is
    suppressed so that the superconducting transition does not take
    place. The inset shows $\beta G_{A,A}(\beta/2)$ at half-filling
    $\mu=0$ at different interaction strengths. The results for the
    $A$ and $C$ sites are identical due to the symmetry of the Lieb
    lattice.}\label{fig.susc} 
\end{figure}

\begin{figure}
  \centering
  \includegraphics[width=\columnwidth]{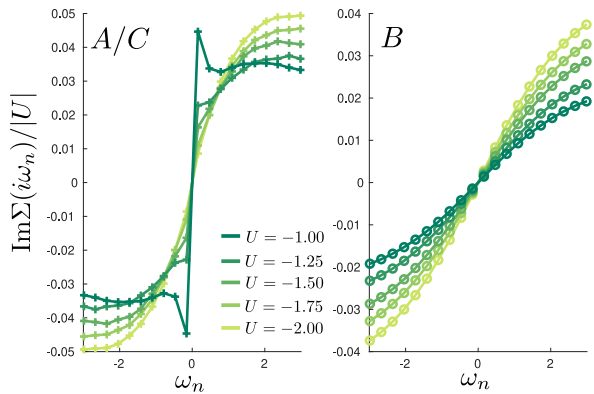}
  \caption{Imaginary part of the self-energy at half-filling and
    inverse temperature $\beta=20$. The superconducting order is
    suppressed. The left panel shows the divergence of the
    self-energy, a signature of a non-Fermi liquid insulator,
    at low interactions on the $A/C$ sites. The right panel shows the
    behavior at the $B$ site, which is the one expected for a Fermi
    liquid.} \label{fig.ses}
\end{figure}

We first study the local contributions to the static spin
susceptibility, given by   
\begin{equation}
  \chi_{\alpha}^{\rm spin} = \frac{1}{\beta^2}
  \sum_{\omega,\omega'}\left(\chi^{{\rm
      ph},\omega,\omega',\nu=0}_{\up\alpha,\up\alpha,\up\alpha,\up\alpha}
  - \chi^{{\rm
      ph},\omega,\omega',\nu=0}_{\up\alpha,\up\alpha,\down\alpha,\down\alpha}\right)
  .
\end{equation}
These susceptibilities are shown in
Fig.~\ref{fig.susc}(a) at half filling $\mu=0$. The 
susceptibility at the $B$ site increases monotonously when the
temperature is lowered. This is consistent with Fermi liquid
behavior. On the $A/C$ sites at both $U=-1$ and $U=-2$, the spin
susceptibility increases down to $T\approx 0.2|U|$, at which point it
decreases rapidly. This indicates a pseudogap at the $A/C$ sites. We
find this pseudogap also away from the flat band, as discussed
in~\cite{supplementary}.

To get further information about the nature of the normal state, we
study the Green's functions in the middle of the imaginary time
interval, $G_{\alpha\alpha}(\beta/2)$. This quantity is related to the
orbital-resolved local spectral function $\mathcal{A}_{\alpha}$
by~\cite{Gull2009,Trivedi1995}  
\begin{equation} 
  \beta G_{\alpha\alpha}(\beta/2) = \int \frac{{\rm d}\omega}{2\pi T}
  \frac{\mathcal{A}_{\alpha}(\omega)}{{\rm cosh}[\omega/(2T)]}.
\end{equation}
Since the integral is dominated by the range $\omega\leq T$, $\beta
G_{\alpha\alpha}(\beta/2)$ approximates $\mathcal{A}_i(\omega=0)$ at low
temperatures. The advantage of studying $\beta
G_{\alpha\alpha}(\beta/2)$ is 
that it avoids the analytical continuation necessary to obtain the
spectral function within DMFT. However, since $\beta
G_{\alpha\alpha}(\beta/2)$ 
is only a reasonable approximation for the spectral function at low
temperatures, its study requires a suppression of the superconducting
order so that the phase transition is avoided. Regardless, it provides
us with qualitative information about the nature of the normal state. 

In a Fermi liquid at zero temperature, the spectral function becomes
the orbital resolved non-interacting density of states
$\rho_{\alpha}$, evaluated at chemical potential shifted by the
self-energy $\mathcal{A}_{\alpha}=\rho_{\alpha}(\mu-{\rm Re 
  \Sigma(\omega=0)})$. In the Lieb lattice, the non-interacting
density of states at the $A/C$ sites is infinite at $\mu=0$ due to the
flat band. As can be seen from  Fig.~\ref{fig.susc}(b), this feature
is not present in the spectral function at large interactions. For low
interaction strengths, a peak in $\beta G_{AA}(\beta/2)$ is still
visible, but when the interaction is increased, $\beta
G_{AA}(\beta/2)$ becomes depleted in an increasingly wide region. This
confirms the non-Fermi liquid behavior in the spin susceptibility at
the $A/C$ in Fig.~\ref{fig.susc}(a) is indeed related to a pseudogap
in the normal state.

While the spectral function tells about a pseudogap, at low
interactions, the imaginary part of the self-energy is helpful in
characterizing the flat band normal state. In a Fermi
liquid, ${\rm Im}\Sigma(i\omega_n)$ vanishes linearly at low
frequencies, ${\rm Im}\Sigma(i\omega_n)\approx i\omega_na+b$. 
As shown in
Fig.~\ref{fig.ses},
this is observed at the $B$ site. At a low interaction $U= -1$,
${\rm 
  Im}\Sigma_{A/C}(\omega)$ instead seems to diverge at $\omega=0$. As
the interaction strength is increased, the divergence at $\omega=0$
disappears and the linear behavior expected for a Fermi liquid is
recovered around $U\approx -1.75$. The imaginary part of the
self-energy is related to the quasiparticle weight $Z$ by
\begin{equation}
  Z=\left( 1-\frac{{\rm Im} \Sigma(i\omega_n)}{\omega_n}
  \bigg|_{\omega_n\to 0} \right)^{-1}.
\end{equation}
Due to the momentum-independence of the self-energy within DMFT,
$Z=m/m^*$, where $m$ is the bare mass and $m^*$ is the effective
mass~\cite{Muller1989}.
The divergence of ${\rm Im}\Sigma(i\omega_n)$ around $\omega=0$
thus indicates a divergence of the effective mass. Therefore, at low
interaction strengths, the divergence in the self-energy at zero
frequency indicates insulating behavior.

The results we present for the self-energy are at a low temperature
with the superconducting order suppressed. This is necessary because
the zero frequency is not accessible on the discrete fermionic
Matsubara frequency scale. At high temperatures, we observe the
diverging behavior in the self-energy already at higher interaction
strengths, but this is likely an artifact of the discrete frequency
scale: ${\rm Im}\Sigma(i\omega_n)$ vanishes so fast at $\omega=0$ that
the linear regime is entirely below the lowest Matsubara
frequency. The diverging behavior in the self-energy at $U=-1$ and
below subsists to very low temperatures, indicating an insulating
state as opposed to a metal with very high quasiparticle effective
mass. 

We thus find two different non-Fermi liquid phases in the normal
state. When the hopping amplitude is of the order of the interaction
or larger, the self-energy at the $A/C$ sites diverges at $\omega=0$,
indicating insulating behavior related to the flat band. When the
interaction is increased, the insulating behavior disappears as shown
by Fig.~\ref{fig.ses}(a), and the spectral function is increasingly
suppressed at low temperatures (Fig.~\ref{fig.susc}(a)). This is a
pseudogap phase. The spin susceptibility (Fig.~\ref{fig.susc}(a))
shows non-Fermi liquid features indicating a pseudogap even at
interaction $U=-1$, suggesting the pseudogap and insulator can
coexist. However, the onset temperature of the pseudogap becomes
vanishingly close to the superconducting critical temperature at low
interaction strengths. In both the insulator and pseudogap cases, the
non-Fermi liquid around half-filling is linked to the flat band, and
the behavior at the $B$ site is that of a Fermi liquid.

A pseudogap was also predicted in the normal state of the Lieb lattice
in~\cite{Swain2020}. In this Monte Carlo study, a metallic state is
predicted at low interaction strengths and a pseudogap phase with
short range pairing correlations at intermediate interactions. In
contrast, our results show that the state at low interaction strengths
at the flat band singularity is not a metallic Fermi liquid phase, but
rather an insulator. In agreement with this previous study, we find
that the normal state at interactions above $|U|\approx 1$ is a
pseudogap phase, characterized here by a depletion of the spectral
function and a suppressed spin susceptibility. A similar pseudogap
state was predicted in~\cite{Hofmann2019} for a lattice model with a
quasi-flat band. The onset temperature of the pseudogap in that study
is predicted to be almost proportional to the interaction strength,
which is similar to our result. In summary, the pseudogap is
consistent with previous literature, but the insulator we predict at
low interactions has not been found before.

\begin{figure}
  \centering
  \includegraphics[width=\columnwidth]{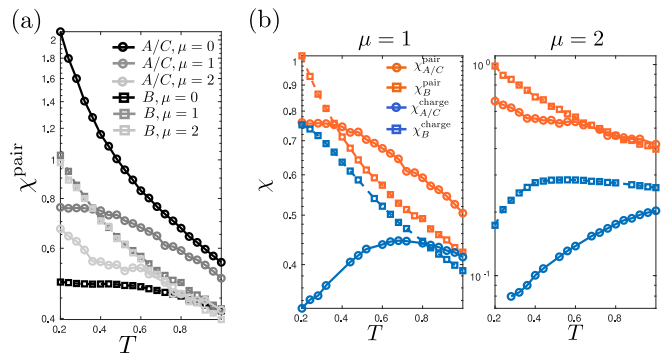}
  \caption{(a) Pairing
    susceptibilities $\chi_{\alpha}^{\rm pair}$ as a function of 
    temperature for different chemical potentials. (b) Pairing and
    charge susceptibilities at $\mu=1$ (left panel) and $\mu=2$ (right
    panel) as a function of temperature. The legend for the right
    panel is the same as for the left. The interaction strength for
    all figures is $U=-2$.} 
  \label{fig.pair}
\end{figure}

In presence of singularities such as van Hove or flat band ones, it is
particularly important to compare competing ordered phases. To this
extent, we study the pairing and charge susceptibilities, 
given by
\begin{align}
  \chi_{\alpha}^{\rm pair} &= \frac{1}{\beta^2}
  \sum_{\omega,\omega'}\chi^{{\rm
      pp},\omega,\omega',\nu=0}_{\up\alpha,\up\alpha,\down\alpha,\down\alpha},
  \\
  \chi_{\alpha}^{\rm charge} &= \frac{1}{\beta^2}
  \sum_{\omega,\omega'}\left(\chi^{{\rm
      ph},\omega,\omega',\nu=0}_{\up\alpha,\up\alpha,\up\alpha,\up\alpha}
  + \chi^{{\rm
      ph},\omega,\omega',\nu=0}_{\up\alpha,\up\alpha,\down\alpha,\down\alpha}\right). 
\end{align}
The pairing susceptibilities at $U=-2$ and different chemical
potentials are shown in Fig.~\ref{fig.pair}(a). At half-filling
$\mu=0$, 
when the flat band is occupied, the pairing susceptibility is strongly
dominated by the $A/C$ sites, where the flat band states reside. On
the other hand, the pairing susceptibility $\chi^{\rm pair}_{\rm B}$
barely increases below $T\approx 0.6$. The total pairing
susceptibility diverges at the critical temperature $T_C\approx 0.11$,
so this shows that the phase transition is driven by pairing at the
$A/C$ sites. When the chemical potential is tuned away from the flat
band, the susceptibility at the $B$ site becomes the dominant
susceptibility. The difference between the susceptibilities in the
different sublattices is however not as pronounced as at the flat
band.

The total pairing susceptibility diverges as the critical temperature
is approached at all chemical potentials. However, even accounting for
the different critical temperatures, the pairing susceptibility at the
flat band singularity is always larger than the local pairing
susceptibilities away from the flat band. The large difference between
the susceptibilities at the $A/C$ and $B$ sites at $\mu=0$ is thus not
only due to a suppression of the susceptibility at the $B$ site, but
the flat band enhances the local pairing susceptibility at the $A/C$
sites. 

An important question is whether the charge susceptibility overtakes
the pairing susceptibility. At half-filling, the pairing and charge
susceptibilities are equal due to the symmetry of the system. Charge
susceptibilities for $\mu=1$ and $\mu=2$ are shown in 
Fig.~\ref{fig.pair}(b). At $\mu=1$, the charge susceptibility at the
$A/C$ sites is suppressed, but $\chi^{\rm charge}_B$ quite close to
the pairing susceptibility on the $B$ site. At low temperatures,
however, the pairing susceptibility grows faster than the charge
susceptibility, so superconductivity is the leading instability. At
$\mu=2$, the charge susceptibility is suppressed at all lattice
sites. It should be noted that the Van Hove singularity in the Lieb
lattice is not at half filling, contrary to the square lattice, hence
the pairing and charge susceptibilities are not equal at the Van Hove
singularity. 

In summary, we found a crossover between two non-Fermi liquid normal
states in the Lieb lattice: an insulator at the $A/C$ sites below
interactions of $|U| \simeq 1$ and a metallic state featuring a
pseudogap above. The insulator and pseudogap can coexist in a small
temperature range. We confirmed that pairing is the leading
instability. The crossover can be observed in present-day ultracold
gas setups since cooling below the critical temperature, often an
obstacle, is not needed, and susceptibilities~\cite{Meineke2012} and
pseudogaps~\cite{Gaebler2010,Feld2011} 
can be measured. In 
twisted bilayer graphene (TBG) the interaction strength is not known,
but mean-field studies indicate it is in the regime where flat band
effects are
significant~\cite{julku:2020,hu:2019,xie:2019,Classen2020}. Thus the
possibility of a flat-band insulator interacting normal state should
be considered in addition to pseudogap~\cite{Jiang2019} and other
exotic 
normal states~\cite{Zondiner2020,Wong2020} already observed,
in particular for other moir\'e materials with stronger flat band
character than TBG. Due to the particle-hole symmetry, our results are
also relevant for flat band
magnetism~\cite{Mielke1992,Tasaki1992,Mielke1993}: the spin
susceptibility maps to the charge susceptibility and {\it vice
  versa}. 

Pseudogap phases have been predicted and observed in many strongly
interacting dispersive
systems~\cite{Huscroft2001,Keller2001,Gull2013,Feld2011}. For instance
in the square lattice with attractive interactions it appears for
large interactions while the weak interaction regime is a
Fermi-liquid~\cite{Keller2001}. Here we showed that a flat band
enhances the pseudogap formation. The insulator behavior found at
small interactions is qualitatively unique to flat
bands. Insulator-superconductor phase transitions are
ubiquitous, controlled by for instance the magnetic
field~\cite{Yazdani1995,Hadacek2004,Baturina2007},
disorder~\cite{Crane2007,Sacepe2011,Lee2015}, or doping
\cite{Semba2001,Konstantinovic2001,Oh2006}. Here, the
insulator-superconductor transition originates from geometry-induced
localization of single particles in a flat band.

{\em Acknowledgments---}
We thank Pramod Kumar for useful discussions.
We acknowledge support by the Academy of Finland under project numbers
303351, 307419, and 327293. Kukka-Emilia Huhtinen acknowledges
financial support by the Magnus Ehrnrooth Foundation. Computing
resources were provided by CSC -- the Finnish IT Centre for Science.

\bibliography{sources}

\end{document}


\title{Supplementary material: \\ Insulator-pseudogap crossover in the Lieb lattice} 
\author{Kukka-Emilia Huhtinen}
\author{P\"{a}ivi T\"{o}rm\"{a}}
\email{paivi.torma@aalto.fi}

\affiliation{Department of Applied Physics, Aalto University, 00076
  Aalto, Finland}

\maketitle

\section{Computation of two-particle quantities within DMFT}

The two-particle Green's function is defined as
\eq{
  G^{(4),{\rm ph}}_{ijkl}(\tau_1,\tau_2,\tau_3) =
  \ave{T_{\tau}[\crea{c}_{i}(\tau_1) \ani{c}_{j}(\tau_2)
      \crea{c}_{k}(\tau_3) \ani{c}_{l}(0)]}.
}
Here, $T_{\tau}$ is the imaginary time ordering operator and $\tau_i$
are imaginary times. The indices $i,j,k,l$ contain the spin and the
orbital $A$, $B$ or $C$. Of
particular interest is the generalized susceptibility
\eq{
  \chi_{ijkl}(\tau_1,\tau_2,\tau_3) =
  G^{(4)}_{ijkl}(\tau_1,\tau_2,\tau_3) -
  G_{ij}(\tau_1,\tau_2)G_{kl}(\tau_3,0).
}
We define the Fourier transform of the generalized susceptibility in
the particle-hole (ph) and particle-particle (pp) channels as follows:
\eqa{
  \chi_{ijkl}^{{\rm ph},\omega,\omega',\nu} =&
  \int_0^{\beta}\int_0^{\beta}\int_0^{\beta} {\rm d}\tau_1{\rm
    d}\tau_2{\rm d}\tau_3 \nonumber\\
  &\chi_{ijkl}(\tau_1,\tau_2,\tau_3)
  e^{-i\omega\tau_1} e^{i(\nu+\omega)\tau_2}
  e^{-i(\nu+\omega')\tau_3},\\
  \chi_{ijkl}^{{\rm pp},\omega,\omega',\nu} =&
  \int_0^{\beta}\int_0^{\beta}\int_0^{\beta} {\rm d}\tau_1{\rm
    d}\tau_2{\rm d}\tau_3 \nonumber\\
  &\chi_{ijkl}(\tau_1,\tau_2,\tau_3)
  e^{-i\omega\tau_1} e^{i(\nu-\omega')\tau_2} e^{-i(\nu-\omega)\tau_3}.
}
Both $\chi^{\rm pp}$ and $\chi^{\rm ph}$ contain the same information,
but these definitions allow to conveniently define the charge and
pairing susceptibilities.

The generalized susceptibilities can be solved with the impurity
solver for the cluster. However, within DMFT, the impurity
susceptibilities are not equal to the lattice
susceptibilities. Instead, the self-consistency is only at the level
of the local irreducible vertex function $\Gamma$. The irreducible
vertex is the two-particle equivalent to the self-energy. The
self-energy is the sum of all irreducible diagrams connecting two
external legs. The irreducible vertex is the sum of all irreducible
diagrams connecting four external legs. However, the reducibility of a
graph depends on the channel, as there are several possible ways to
disconnect pairs of outer legs. Thus the irreducible vertex is
channel-dependent. 

The irreducible vertex is related to the generalized susceptibility by
the Bethe-Salpeter equation~:
\eqa{
  &\chi^{{\rm c},\omega,\omega',\nu}_{ijkl}(\vec{k},\vec{k'},\vec{q}) =
  \chi_{0,ijkl}^{{\rm 
      c},\omega,\omega',\nu}(\vec{k},\vec{k'},\vec{q}) \nonumber \\
  +&
  \chi_{0,ij'ji'}^{{\rm 
      c},\omega,\omega'',\nu}(\vec{k},\vec{k}_1,\vec{q})\Gamma_{i'j'k'l'}^{{\rm 
      c},\omega'',\omega''',\nu}(\vec{k}_1,\vec{k}_2,\vec{q})\chi_{k'kl'l}^{{\rm
      c},\omega''',\omega',\nu} (\vec{k}_2,\vec{k'},\vec{q}). 
}
where repeated indices are summed over and ${\rm c}$ denotes the
channel. Here $\chi_0^{{\rm c}}$ is the
bare contribution to the susceptibility, for example in the
particle-hole channel $\chi_{0,ijkl}^{{\rm ph},\omega,\omega',\nu}
(\vec{k},\vec{k'},\vec{q})= -\beta\delta_{\omega,\omega'}\delta_{\vec{k},\vec{k'}}
G_{il}(\omega,\vec{k})G_{kj}(\omega+\nu,\vec{k}+\vec{q})$. 

\begin{figure}
  \centering
  \includegraphics[width=0.2\textwidth]{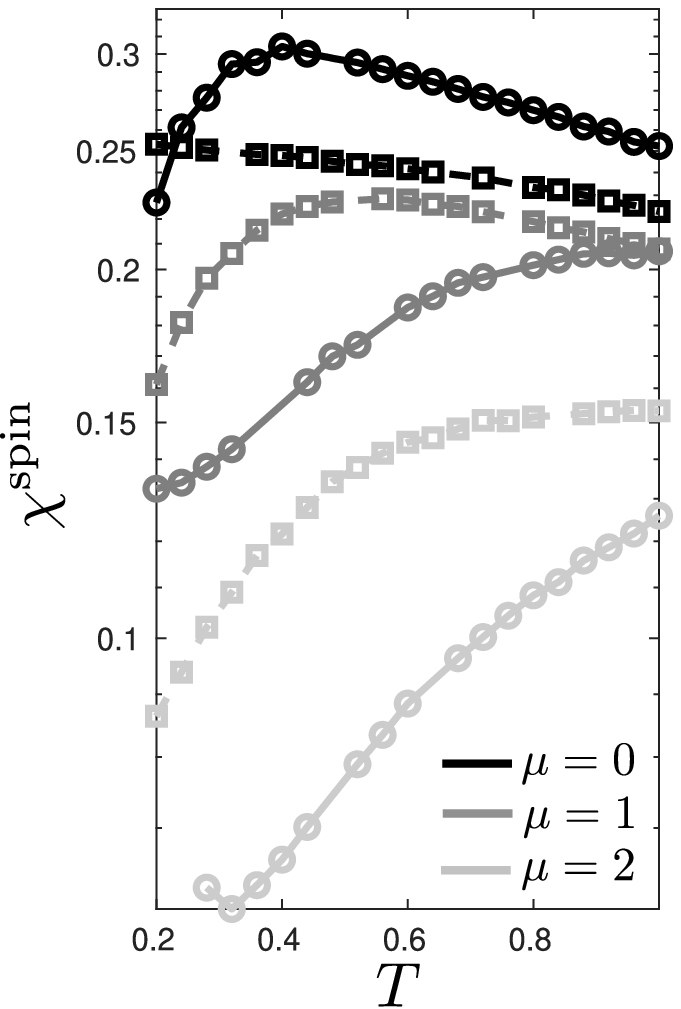}
  \caption{Spin susceptibility as a function of temperature for
    different chemical potentials. The interaction strength is
    $U=-2$. The susceptibility at the $A/C$ sites is plotted with
    dots, and the susceptibility at the $B$ site is plotted with
    squares.}\label{fig.susc}
\end{figure}

In DMFT, the vertex $\Gamma(\vec{q},\vec{k},\vec{k'})$ is approximated
by the local vertex $\Gamma$ of the impurity. Summing over the
momenta $\vec{k}$ and $\vec{k'}$, we obtain
\eqa{
  &\chi^{{\rm c},\omega,\omega',\nu}_{ijkl}(\vec{q}) =
  \chi_{0,ijkl}^{{\rm c},\omega,\omega',\nu}(\vec{q})\nonumber\\
  &+
  \chi_{0,ij'ji'}^{{\rm 
      c},\omega,\omega'',\nu}(\vec{q})\Gamma_{i'j'k'l'}^{{\rm 
      c},\omega'',\omega''',\nu}\chi_{k'kl'l}^{{\rm
      c},\omega''',\omega',\nu} (\vec{q}).
  }

This equation can be written separately for the lattice quantities and
the impurity quantities $\chi_{\rm imp}$ and $\chi_{0,{\rm imp}}$. For
the impurity, we further approximate the susceptibilities by the local
ones, so that
\eq{
  \chi^{{\rm c},\omega,\omega',\nu}_{{\rm imp},ijkl} =
  \chi_{0,{\rm imp},ijkl}^{{\rm c},\omega,\omega',\nu} 
  +
  \chi_{0,{\rm imp},ij'ji'}^{{\rm 
      c},\omega,\omega'',\nu}\Gamma_{i'j'k'l'}^{{\rm 
      c},\omega'',\omega''',\nu}\chi_{{\rm imp},k'kl'l}^{{\rm
      c},\omega''',\omega',\nu}.
}
The local vertex is computed by inverting this equation. Since within
DMFT it is the same as the vertex of the lattice, the lattice
susceptibility can be computed by substituting the result in the
Bethe-Salpeter equation for the lattice model. In matrix form, the
result can be written
\eq{
  [\chi(i\nu,\vec{q})]^{-1} = [\chi_0(i\nu,\vec{q})]^{-1} - [\chi_{0,{\rm
      imp}}(i\nu)]^{-1} + [\chi_{{\rm imp}}(i\nu)]^{-1}. 
}
Here the matrices representing the susceptibilities at each $i\nu$ are
block matrices in the two fermionic Matsubara frequencies $\omega$, $\omega'$.

\section{Pseudogap away from half-filling}

Figure~\ref{fig.susc} shows the local spin susceptibilities at
different chemical potentials. The non-Fermi liquid behavior found
only at the $A/C$ sites on the flat band is found at all sites when
the chemical potential is tuned away from the flat band. A pseudogap
phase thus occurs also away from the flat band, but a qualitative
difference in the behavior at the $A/C$ sites and the $B$ site is
only found at the flat band.